\documentclass[10pt]{article}
\pdfoutput=1
\usepackage{times}
\usepackage{color}
\usepackage{caption}
\usepackage{pgfplots}
\usepackage{todonotes}
\usepackage{cite} 
\usepackage{amsthm}
\usepackage{amsmath}
\usepackage{dsfont}
\usepackage{graphicx}
\usepackage{psfrag}
\usepackage{siunitx}

\newcommand{\N}{\mathds{N}}

\newcommand{\E}{\mathsf{E}}
\newcommand{\To}{\mathop{\rightarrow}}
\newcommand{\set}[1]{\left\{#1\right\}}

\theoremstyle{plain}
\newtheorem{thm}{Theorem}[section]

\newtheorem{lem}[thm]{Lemma}

\newtheorem{exa}[thm]{Example}

\date{}
\begin{document}
\pagenumbering{gobble}

\title{\textbf{\Large Error Propagation Through a Network With Non-Uniform Failure}}

\author{
Sandra K\"{o}nig\\ Digital Safety \& Security Department\\ Austrian Institute of Technology\\ Klagenfurt, Austria\\
sandra.koenig@ait.ac.at}

\maketitle

\begin{abstract}
A central concern of network operators is to estimate the probability of an
incident that affects a significant part and thus may yield to a breakdown.
We answer this question by modeling how a failure of either a node or an
edge will affect the rest of the network using percolation theory. Our
model is general in the sense that it only needs two inputs: the topology
of the network and the chances of failure of its components. These chances
may vary to represent different types of edges having different tendencies
to fail. We illustrate the approach by an example, for which we can even
obtain closed form expressions for the likelihood of an outbreak remaining
bounded or spreading unlimitedly.
\end{abstract}

\section{Introduction}
Information and communication technology (ICT) infrastructures as well as utility
networks (such as water supply, etc.) are commonly modeled as graphs,
representing networks of interconnected components. Failure of one or more
sensitive components may cause an essential damage, while a breakdown of
other components may hardly affect the network performance or functionality.
To estimate the risk of a breakdown of the entire network, it appears
informative to investigate the impact of an arbitrary but known failure
scenario that involves the breakdown of a component (network node) or an
interconnection (network edge). The impact to the overall infrastructure is
determined by the way in which the problem is ``transmitted'' from where it
occurred to other parts of the network. The particular form of such
error propagation may be determined by malfunctions that cause further
malfunctions as well as information (e.g., notifications) that spread in the
network. The process can therefore be viewed similarly to the spread of an
epidemic \cite{Poggi2013}. This perspective is particularly adopted in
economics \cite{Duffie2009,Duffie2011}, however and somewhat surprisingly, 
it has not yet seen much application in risk management. In this work, we take
first steps to close this gap.

To understand how certain events trigger other events in a network,
percolation theory \cite{Grimmett1989} has evolved into an indispensable tool. Especially models
of disease spreading \cite{Newman2002,Poggi2013} in terms of percolation have
become popular over the last decades. Most of these models are built for a
specific class of networks such as scale-free networks \cite{Schwartz2002},
\cite{Cohen2002} or lattices \cite{Sander2002} (one treatment of a general
networks can be found in \cite{Miller2008}). Moreover, it
is a common assumption that all contacts are equally likely to transmit the disease.
While this might be a reasonable simplification to treat disease problems, such assumptions (and
hence the resulting models) are too restrictive to properly model the
situation in ICT and utility networks.
There, components have a much better understood heterogeneity, which strongly affects how an error in one component
can (or cannot) affect other touching components\footnote{While such
heterogenity is obviously also present in social structures (``networks of
humans''), the characteristics of different individuals are highly diverse
and difficult to elicit, which justifies the simplifications made in the past
literature.}.

Hence we here propose a model for error propagation that accounts for more
diversity, in the sense that we can classify and distinguish network
components in terms of (common) properties that are relevant for error
propagation. As a very simple example, we may divide the components in a
utility network into those of ``high significance'', meaning that a failure
may have devastating consequences, and those of ``low significance'', which
could mean that a failure can be instantly covered by a redundant fall-back
subsystem taking over the functionality of the broken component. Practical
classifications may of course be much more complex and sophisticated.

Applying percolation theory to such a setting has been attempted in the past:
In \cite{Restrepo2008}, nodes may fail with different probabilities but the
arguments are merely heuristic. In \cite{Meyers2006}, the standard assumption
has been softened by allowing both directed and undirected contacts, yielding
a so called semi-directed network. In our work we use similar argumentation
based on generating functions but our model differentiates from theirs in the
following fundamental aspects: We deal with a fully directed network as an
undirected edge can be interpreted as a pair of directed edges pointing in
opposite directions. Moreover, we consider different classes of edges having
different probabilities to transmit an error. These classes can be seen as a
characterization of different types of links, i.e. links made of different
materials or links that enjoy different levels of protection.

Under these assumptions, we compute the probability that an error affects a
significant number of components, which will be termed an \emph{epidemic}, as
well as how many nodes are indeed affected in this case. In this context
an epidemic means an unlimited outbreak, or equivalently, an unbounded
propagation of errors.

\subsection{Preliminaries and Notation}\label{sec:notation}
Throughout the paper, we denote vectors by bold printing, and functions by
calligraphic letters. Sets, matrices and random variables appear as uppercase
letters.

The topology of a network can be described by its \emph{degree distribution}
$P(k)$, giving the probability of a node having exactly $k$ neighbors. A
directed network is analogously characterized by the joint probabilities
$P(j,k)$ that a node has exactly $j$ ingoing and $k$ outgoing edges.
Furthermore, we will need to consider the \emph{excess degree distribution},
usually denoted by $Q(k)$ or $Q(j,k)$, respectively, which describes the following situation: if we
arrive at a random node via an edge, what is the distribution of the
remaining degree, i.e. how many edges remain when we exclude the one we
traveled along? Due to the biased situation (nodes with higher degree are more likely to be reached) 
 this distribution is not just a reduced version of the original degree but is going to be calculated in the next section. 

Our main focus will be on ``occupied`` edges, denoting those edges that are
affected by the failure under consideration. Together with the nodes incident
to them, they build up the so called ``occupied`` network, as opposed to the
``original network'' in which no incident has yet occurred. In what follows,
we add the superscript ``o'' (for ``original'') to quantities corresponding
to the network state before a failure. Variables describing the network after
failure (the occupied state) have no such superscript.

Working with probability distributions directly is possible, yet it turns out
that a different representation greatly simplifies matters here: each
discrete probability distribution $P(k)$ of a random variable with
realizations $k\in\N$ can conveniently be represented by a so called
\emph{generating function}, defined for the distribution $P$ as,
\[
\mathcal{G}(x):=\sum_{k=0}^{\infty}P(k)x^k.
\]
These functions are easier to handle then the distributions themselves and
are therefore applied in many different fields \cite{Wilf1994}, including the analysis of networks (see
e.g. \cite{Newman2001} for an overview). Generating functions exhibit many
nice properties, one of which comes most handy for our analysis here:
\begin{lem}\label{lem:generating-function-power}
If the
function $\mathcal{G}$ generates the degree distribution of a node, then the
power $\mathcal{G}^m$ generates the distribution of the total degree of $m$
independent nodes.
\end{lem}
In the following, we will denote the generating function of a (ordinary)
degree distribution by $\mathcal{G}$ and the corresponding generating function of the excess degree
distribution by  $\mathcal{H}$.

\section{Modeling Non-Uniform Failures}
Let a physical network be modeled as a graph $G(V,E)$, where $V$
is a finite set of nodes and $E$ is the set of edges between them. Since we
consider error propagations being determined by the different nature of interaction between nodes,
 we partition the full edge set $E$ into $n$ \emph{classes} of
edges. Hereafter, we say that an edge is ``of type $i$'', to distinguish it
from other edges with different properties. The index $i$ thus ranges over
$n$ classes of edges, rather than the physical edges themselves. Edges of the
same type are indistinguishable, while edges of different types have
distinctive properties related to the error propagation.

Notice that this partitioning of edges does not need to be extended to nodes
as well. To see this, simply consider a problem that reaches node $v\in V$
over an edge $e\in E$. If $v$ reacts robustly in the sense of blocking the
error from further spreading, then this is equal to assume that $e$ already
blocked the error. Likewise, if $v$ fails and the error propagates further,
then this effect becomes visible by the other edges transporting the problem
to their end points, and so on.
If only a node fails but the error is not
further propagated, then we can model this effect by thinking of the node
with an internal edge that can bound the error within the node and prevents
its further propagation to outgoing nodes from this edge (this is a standard
modeling trick that is, for example, also applied to compute maximal flows in
graphs with vertex rather than edge capacities).

When investigating communication networks under random failures, it has
been a common assumption that edges fail (in the sense of forwarding an error from node $i$ to its neighbor)
 independently  of each other  and with uniform probability $p$ \cite{Cohen2000,Cohen2001, Callaway2000, Newman2011, Newman2013}.
Using the formalism introduced in Section \ref{sec:notation},
the  network of occupied edges is then described by the degree distribution
\begin{equation*}
P(j,k)=\sum_{j^o, k^o} P^o(j^o, k^o) \cdot \binom{j^o}{j}p^{j}(1-p)^{j^o-j} \cdot \binom{k^o}{k}p^{k}(1-p)^{k^o-k},
\end{equation*}
where  $P^o$ denotes the original degree distribution before any failure, and
$j^{o}, k^{o}$ denote the number of incoming and outgoing edges in that network, respectively.
In reality, edges are usually \emph{not} equally likely to fail. We account for
this by our partitioning of $E$ into edges of different types
$i\in\set{1,2,\ldots,n}$. Each type $i$ has a specific probability $p_i$ to
fail, or be ``occupied`` as we denote it here (see Section \ref{sec:notation} again), and its breakdown
is assumed to be independent of that of any other edge. In this
case we have to replace the single binomial distribution by a product of
binomial distributions and the degree distribution of the occupied network
$$P(\textbf{j},\textbf{k}):=P(j_1, \ldots, j_n;k_1, \ldots, k_n)$$ with
vectors $\textbf{j}=(j_1, \ldots, j_n)$ and $\textbf{k}=(k_1, \ldots, k_n)$
computes as

\begin{equation}\label{eq:dist-after-failure}
P(\textbf{j},\textbf{k})=\sum_{\textbf{j}^o \geq \textbf{j}}\sum_{\textbf{k}^o \geq \textbf{k}} P^o(\textbf{j}^o,\textbf{k}^o) \prod_{i=1}^n \binom{j_i^o}{j_i}p_i^{j_i}(1-p_i)^{j_i^o-j_i} \cdot \prod_{i=1}^n \binom{k_i^o}{k_i}p_i^{k_i}(1-p_i)^{k_i^o-k_i}
\end{equation}

where $\textbf{j}^o \geq \textbf{j}$ is a shorthand for the component-wise
inequalities $j_i^o \geq j_i$ for all $i$. Similarly as before, $j_i$ and $k_i$ denote the number of incoming and outgoing edges of type $i$ and $p_i$ is the probability that an edge of type $i$ conveys an error. We assume this probability to be only dependent on the type, hence incoming and outgoing edges of the same type have equal probability to fail.

Starting from a different point of view, a similar model has been proposed
by \cite{Meyers2006} for semi-directed networks. Still, our models are different in at least two ways:
by considering an undirected edge as a pair of one incoming and one outgoing edge, we restrict our
attention to fully directed networks. Further, we do not limit ourselves to
two types of edges but allow for any finite number $n$ of classes.

Knowing the degree distribution after random failure of edges, statements
about expected size of an outbreak or the probability of an epidemic can be
made with help of generating functions as introduced in Section \ref{sec:notation}.
Let the generating function for the original distribution $P^o$ be
$\mathcal{G}^o$, which is
\[
\mathcal{G}^o(x_1, \ldots, x_n; y_1, \ldots, y_n)=\sum_{\textbf{j}^o,\textbf{k}^o\geq \textbf{0}} P^o(\textbf{j}^o,\textbf{k}^o)\prod_{i=1}^n x_i^{j_i^o} y_i^{k_i^o}
\]
As average in- and out-degrees have to match for all types of edges, we get
the constraints
\[
\frac{\partial}{\partial x_i} \mathcal{G}^o(1, \ldots, 1; 1, \ldots, 1)=\frac{\partial}{\partial y_i} \mathcal{G}^o(1, \ldots, 1; 1, \ldots, 1)=: z_i^o
\]
for  all $i$. Furthermore we will use generating functions $\mathcal{H}_i^o$
for the excess degree distributions. These are especially handy when dealing
with occupied edges (see Section \ref{sec:notation}), since they relate to
the possible paths (edges) along which an error can further propagate after
it has reached a node along an edge of type $i$. To find the generating
function of these excess degree distribution we note that it is more likely
to arrive at a random node with higher degree $j_i$ of incoming edges. Once
an edge has transmitted an error the corresponding degree reduces by one,
thus we get (by direct calculation) a relation to the generating function
of the original network by
\begin{align*}
&\mathcal{H}_i^o(x_1, \ldots, x_n; y_1, \ldots, y_n)\\
&=\frac{\sum_{\textbf{j}^o,\textbf{k}^o\geq \textbf{0}}j_i^o  P^o(\textbf{j}^o,\textbf{k}^o)x_1^{j_1^o}\ldots x_{i-1}^{j_{i-1}^o}x_i^{j_i^o -1}x_{i+1}^{j_{i+1}^o}\ldots x_n^{j_n^o} \cdot y_1^{k_1^o}\ldots y_n^{k_n^o}}{\sum_{\textbf{j}^o,\textbf{k}^o\geq \textbf{0}}j_i^o  P^o(\textbf{j}^o,\textbf{k}^o)}\\
&=\left[\frac{\partial}{\partial x_i}  \mathcal{G}^o(x_1, \ldots, x_n; y_1, \ldots, y_n)\right] / z_i^o.
\end{align*}

On the other hand, the generating function for the degree distribution of \emph{occupied} edges is defined by
\[
\mathcal{G}(x_1, \ldots, x_n; y_1, \ldots, y_n)=\sum_{\textbf{j},\textbf{k}\geq \textbf{0}} P(\textbf{j},\textbf{k})\prod_{i=1}^n x_i^{j_i} y_i^{k_i}
\]
where $P(\textbf{j},\textbf{k})$ is given in \eqref{eq:dist-after-failure}.
This function can be expressed in terms of the generating function of the
original distribution by straightforward calculation
\begin{align*}
&\mathcal{G}(x_1, \ldots, x_n; y_1, \ldots, y_n)\\
&=\sum_{\textbf{j},\textbf{k}\geq \textbf{0}} \sum_{\textbf{j}^o \geq \textbf{j}}\sum_{\textbf{k}^o \geq \textbf{k}}\left[P^o(\textbf{j}^o,\textbf{k}^o) \prod_{i=1}^n  \binom{j_i^o}{j_i}p_i^{j_i}(1-p_i)^{j_i^o-j_i}\right.\\
&\qquad\qquad\qquad\qquad\qquad\qquad\times\left.\prod_{i=1}^n  \binom{k_i^o}{k_i}p_i^{k_i}(1-p_i)^{k_i^o-k_i} \prod_{i=1}^n x_i^{j_i} y_i^{k_i}\right]\\
&=\sum_{\textbf{j}^o, \textbf{k}^o \geq \textbf{0}} P^o(\textbf{j}^o,\textbf{k}^o) \prod_{i=1}^n \left[ \left(\sum_{j_i=0}^{j_i^o} \binom{j_i^o}{j_i}p_i^{j_i}(1-p_i)^{j_i^o-j_i}x_i^{j_i}\right)\right.\\
&\qquad\qquad\qquad\qquad\qquad\qquad\left.\times\left(\sum_{k_i=0}^{k_i^o} \binom{k_i^o}{k_i}p_i^{k_i}(1-p_i)^{k_i^o-k_i} y_i^{k_i}\right) \right]\\
&=\sum_{\textbf{j}^o, \textbf{k}^o \geq \textbf{0}} P^o(\textbf{j}^o,\textbf{k}^o) \prod_{i=1}^n \left[\sum_{j_i=0}^{j_i^o}\binom{j_i^o}{j_i}(x_i p_i)^{j_i}(1-p_i)^{j_i^o-j_i} \right]\\
&\qquad\qquad\qquad\qquad\qquad\qquad\times\left[\sum_{k_i=0}^{k_i^o}\binom{k_i^o}{k_i}(y_i p_i)^{k_i}(1-p_i)^{k_i^o-k_i} \right] \\
&=\sum_{\textbf{j}^o, \textbf{k}^o \geq \textbf{0}} P^o(\textbf{j}^o,\textbf{k}^o) \prod_{i=1}^n \left[x_ip_i + (1-p_i) \right]^{j_i^o}\left[y_ip_i + (1-p_i) \right]^{k_i^o}\\
&=\mathcal{G}^o\Big((1-p_1+x_1p_1), \ldots, (1-p_n+x_np_n);\\
&\qquad\qquad\qquad\qquad\qquad\qquad (1-p_1+y_1p_1), \ldots, (1-p_n+y_np_n)\Big)
\end{align*}
with use of the binomial theorem. Similarly, one finds a relation between the
generating functions for excess degrees for all $i$:
\begin{align*}
&\mathcal{H}_i(x_1, \ldots, x_n; y_1, \ldots, y_n)=\\
&\mathcal{H}_i^o\Big((1-p_1+x_1p_1), \ldots, (1-p_n+x_np_n);
(1-p_1+y_1p_1), \ldots, (1-p_n+y_np_n)\Big),
\end{align*}
as the average degrees of occupied edges of type $i$ are $z_i=z_i^o p_i$.

\section{Predictions for Small Scale Outbreaks}\label{sec:finite}
Based on these generating functions, we can now predict the influence of an
error spreading over the network. Especially, we are interested in the
average number of nodes affected by an error starting to spread from a given
node as well as the probability that it affects an extensive number of nodes
(i.e. it causes an epidemic).

Let $H_i$ denote the generating functions for the size $w$ of an outbreak (measured in the number of nodes affected) arising from
 failure of an edge of type $i$.
A relation to the degree distributions of occupied edges can be found based
on the following decomposition. Assume we arrive at a node with occupied
degree $(j_1, \ldots, j_n; k_1, \ldots, k_n)$, then each of the $k_i$ edges
of type $i$ may convey the error and the corresponding probability is by
Lemma \ref{lem:generating-function-power} described by the function
$(H_i(w))^{k_i}$.
Thus we get
\begin{align*}
H_i(w)&=w\frac{\sum_{\textbf{j}\geq\textbf{0}} j_i \left[P(\textbf{j},\textbf{0})+\sum_{\textbf{k}: k_1+\ldots+k_n=1}P(\textbf{j},\textbf{k})\prod_{i=1}^n \left(H_i(w)\right)^{k_i} + \ldots \right]}{\sum_{\textbf{j},\textbf{k}\geq \textbf{0}}j_i  P(\textbf{j},\textbf{k})} \\
&=w \frac{\sum_{\textbf{j}\geq\textbf{0}}\sum_{\textbf{k}\geq\textbf{0}} j_i P(\textbf{j},\textbf{k})\prod_{i=1}^n \left[H_i(w)\right]^{k_i}}{z_i} \\
&=w \cdot \mathcal{H}_i(1, \ldots, 1;  H_1(w), \ldots, H_{n}(w)),
\end{align*}
for all edge types $i$. Correspondingly, one computes the generating function $G$ for the distribution of outbreak sizes originating from an infected node as
\[
G(w)=w \cdot \mathcal{G}(1, \ldots, 1; H_1(w), \ldots, H_{n}(w)).
\]
With the help of these equations, we can now set out to compute the average
size of an outbreak resulting from any possible failure.

If an edge of type $i$ failed we have to solve the linear equation system
\begin{equation}\label{eq:system-Hi}
H_i^{\prime}(w)=1+\sum_{j=1}^n \frac{\partial}{\partial y_j} \mathcal{H}_i(1, \ldots, 1; H_1(w), \ldots, H_{n}(w)) \cdot H_j^{\prime} (w)
\end{equation}
to find the expected number $S_i$ of affected nodes
\[
\E[S_i]=H_i^{\prime}(1).
\]
Similarly, the expected size of an outbreak starting from a random node is
given by
\begin{equation}\label{eq:finite-size}
\E[S]=G^{\prime}(1)
=1+\sum_{j=1}^n \frac{\partial}{\partial y_j} \mathcal{G}(1, \ldots, 1; H_1(w), \ldots, H_{n}(w)) \cdot H_j^{\prime} (w) \bigg\vert_{w=1}
\end{equation}
for the case of finite outbreaks (the fraction of nodes affected by an
epidemic are determined later).

In both cases, we need to solve a linear system and hence adapt the usual
notation $Ax=b$ to describe it. The system \eqref{eq:system-Hi} is than
determined by $b=(1, \ldots, 1)^T$ and the coefficient matrix with
\[
a_{ij}=
- \frac{\partial}{\partial y_j}\mathcal{H}_i (1, \ldots, 1; 1, \ldots, 1)= - \frac{\partial}{\partial y_j}\mathcal{H}^o_i (1, \ldots, 1; 1, \ldots, 1)\cdot p_j
\]
for $i\neq j$ and
\[
a_{ii}=1-\frac{\partial}{\partial y_i}\mathcal{H}_i (1, \ldots, 1; 1, \ldots, 1)=1-\frac{\partial}{\partial y_i}\mathcal{H}^o_i (1, \ldots, 1; 1, \ldots, 1)\cdot p_i
\]
on the diagonal. This system can be solved numerically and the solution can then be plugged into equation \eqref{eq:finite-size}.

\begin{exa}\label{exa:ER}
If we adapt the well known Erd\H{o}s-R\'enyi model \cite{Erdos1959} for all
edges with probabilities $q_i$ for existence of en edge of type $i$, we get
the generating function for the joint degree distribution before failure
\begin{align*}
\mathcal{G}^o(x_1, \ldots, x_n; y_1, \ldots, y_n)&=\sum_{\textbf{j},\textbf{k}\geq \textbf{0}} \prod_{i=1}^n \frac{(nq_i)^{j_i}}{j_i!}e^{-nq_i}\frac{(nq_i)^{k_i}}{k_i!}e^{-nq_i} x_i^{j_i}y_i^{k_i}\\
&=\prod_{i=1}^n e^{nq_i(x_i+y_i-2)}=e^{n\sum_{i=1}^n q_i(x_i +y_i-2)}
\end{align*}
which agrees with the generating function $\mathcal{H}_i$ for the excess
degree distribution since $z_i=nq_i$.
As before, we denote by $p_i$ the probability that an edge of type $i$ 
fails. Then, the matrix of the system to be solved has entries
\[
a_{ij}=\begin{cases}
- np_jq_j &\mbox{if } i\neq j\\
1-np_jq_j & \mbox{if } i=j
\end{cases}
\]
for $p_j, q_j \in [0,1]$. The determinant of this matrix $A$ can be computed
for example via induction on $n$ and turns out to be $1-np_1q_1 - \ldots
-np_nq_n$. Note, however, that neither the $p_i$'s nor the $q_i$'s do have to
sum to one as they do \emph{not} describe a probability distribution but
rather denote different parameters. At the same time, $p_iq_i$ can be thought
of as the probability for existence of an edge of type $i$ transporting an error in the network, thus belonging to the `infected' part of the network after failure. 

The resulting equation system is easy to solve for this specific matrix: a
direct computation shows that $(1, \ldots. 1)^T$ is an eigenvector with
eigenvalue $\det(A)$, hence $x=1/\det(A)\cdot(1, \ldots, 1)^T$ is the sought
solution for $(H_1^{\prime}(1), \ldots, H_n^{\prime}(1))^T$. Thus, the
expected size of an outbreak after failure of an edge of type $i$ is
\[
\E[S_i]=\frac{1}{\det(A)}=\frac{1}{1-np_1q_1 - \ldots -np_nq_n}
\]
Substituting our solution into \eqref{eq:finite-size} yields the same
expected size of an outbreak resulting from failure of a node:
\[
\E[S]=\frac{1}{1-np_1q_1 - \ldots -np_nq_n}=\E[S_i]
\]
for all $i$. So the critical threshold corresponds to the situation where the
terms $p_iq_i$ sum to $1/n$ (i.e. the denominator is not zero).
\end{exa}

The solution of these systems always contains a factor $\frac{1}{\det(A)}$
from inversion of $A$ and hence the expected size of an outbreak will become
extensive if this determinant vanishes. From the identity
\[
\frac{\partial}{\partial y_j}\mathcal{H}_i(x_1, \ldots, x_n: y_1, \ldots, y_n)=\frac{\partial}{\partial y_j}\mathcal{H}^o_i(x_1, \ldots, x_n: y_1, \ldots, y_n)\cdot p_j
\]
for all $i,j \in \{1, \ldots, n\}$, we get a condition on the probabilities
$p_i$ that represents the beginning of a possible epidemic. On the other hand
we can say that an epidemic is avoided, if the determinant is strictly
positive.

\begin{exa}\label{exa:ER-condition}
Under the assumptions of Example \ref{exa:ER}, one finds that if the
criterion
\begin{equation}\label{eqn:finite-outbreak-criterion}
1-np_1q_1 - \ldots -np_nq_n>0
\end{equation}
is fulfilled, then no epidemic will occur as the expected number of affected nodes is finite.
\end{exa}

\section{Predictions for Large Scale Outbreaks}
Let us assume that an epidemic is possible, e.g. by a violation of the condition from Example
\ref{exa:ER-condition}. Generally, we are interested in answering two
questions: First, what is the probability $P_{ep}$ of an epidemic, i.e. that
the fraction $f$ of nodes affected scales with the network size? Second, what
fraction of nodes is affected if an epidemic occurs? While formula
\eqref{eq:finite-size} is not longer working due to the increased likelihood
of loops, these two quantities can be identified with quantities known from
percolation theory \cite{Meyers2006}. In order to make this relation explicit
we need some notation: let GSCC denote the ``giant strongly connected
component''. This is a subset of network components, consisting of all nodes
that can be reached by any other node of that component.
The sets GIN and GOUT comprise all nodes that lead into GSCC or can be reached from nodes in
GSCC. Both are themselves not part of GSCC, respectively (more about these
components can for example be found in \cite{Cohen2010}). With these
definitions, one finds that $P_{ep}$ equals the relative size of GSCC and GIN
(in relation to the total number of nodes in the network)
as from those node a giant fraction of nodes can be reached. Similarly, $f$
equals the relative size of GSCC and GOUT together.

A convenient way to compute these quantities is to consider what we call the \emph{dual} network consisting of
the same nodes as the original one but with all edges pointing in opposite
direction. Extending our notation accordingly, we add a superscript ``d'' to
all objects that relate to the dual network. The probability generating
function for the degree distribution of this dual network is then given by
\[
\mathcal{G}^d(x_1, \ldots, x_n; y_1, \ldots, y_n)=\mathcal{G}(y_1, \ldots, y_n; x_1, \ldots, x_n)
\]
and as before we compute the generating function $G^d$ for the number
$w$ of nodes affected by failure of a
random node in the dual network as
\[
G^d(w)=w \cdot \mathcal{G}^d(1, \ldots, 1;  H^d_1(w), \ldots, H^d_{n}(w))=w \cdot \mathcal{G}(H^d_1(w), \ldots, H^d_{n}(w);1, \ldots, 1)
\]
where again $H^d_i$ generate the probability distributions of the number of
nodes ultimately affected by failure of an edge of type $i$.

Then, $G^d(1)$ gives the probability that only finitely many nodes can be
reached in the dual network from a specific node which can be viewed as the
probability that this node is only reached by finitely many nodes in the
original network. Hence (with the calculations above) the probability that it is reached from a giant
fraction of nodes is
\[
f=1-G^d(1)=1- \mathcal{G}(H^d_1(1), \ldots, H^d_{n}(1);1, \ldots, 1),
\]
which corresponds to the fraction of nodes affected in case of an epidemic
caused by failure of a node. However, we stress an important difference to
our calculations in Section \ref{sec:finite}: if an epidemic is possible,
then the functions $H^d_i$ are not normalized in the sense that $H^d(1)$ now
may be strictly smaller than 1, as this only gives the probability for an
outbreak of finite size. Therefore, we first have to solve the (nonlinear)
equation system
\begin{equation}\label{eq:epidemic-system}
H^d_i(1)=\frac{\partial}{\partial y_i}\mathcal{G}(H^d_1(1), \ldots, H^d_n(1); 1, \ldots, 1) / z_i
\end{equation}
for $H^d_i(1)$ for all $i$. This system most likely does not admit an
analytical solution, but numerical approximations can do sufficiently well.
As before, these equations can be rephrased in terms of the generating
functions corresponding to the original network before failure (as this is
the distribution we can measure directly when analyzing the risk of a
network).

The probability of an epidemic can be calculated analogously to or directly
from the results for the case of finite outbreaks. The only difference to our
previous derivations is that we now have to distinguish according to the
number of incoming edges instead of outgoing edges (and hence again use $H_i$
to denote the number of nodes affected rather than $H_i^d$ that corresponds
to the number of nodes from which a node can be reached). It thus follows
that
\[
P_{ep}=1- \mathcal{G}(1, \ldots, 1;H_1(1), \ldots, H_{n}(1)),
\]
where this time $H_i(1)$ are determined by
\[
H_i(1)=\frac{\partial}{\partial x_i}\mathcal{G}(1, \ldots, 1; H_1(1), \ldots, H_n(1)) / z_i
\]
for all $i$. Again, this system may be solved numerically.

In order to get an impression on how to compute such quantities, let us
reconsider Example \ref{exa:ER}.

\begin{exa}
Assume again an Erd\H{o}s-R\'enyi model for all kind of edges so that an edge
of type $i$ exists with probability $q_i$ in the original network before any
failure. In Example \ref{exa:ER} we found
\[
\mathcal{G}^o(x_1, \ldots, x_n; y_1, \ldots, y_n)=\mathcal{H}_i^o(x_1, \ldots, x_n; y_1, \ldots, y_n)=e^{n\sum_{i=1}^n q_i(x_i +y_i-2)}
\]
which yields
\[
\mathcal{G}(x_1, \ldots, x_n; y_1, \ldots, y_n)=\mathcal{H}_i(x_1, \ldots, x_n; y_1, \ldots, y_n)=e^{n\sum_{i=1}^n q_i\cdot p_i(x_i +y_i-2)},
\]
and which can again be viewed as an Erd\H{o}s-R\'enyi model with reduced
probability $q_i p_i$ for existence of an occupied edge of type $i$. Hence, in order
to find the fraction $f$ of affected nodes we need to solve the system
\[
H_i^d(1)=\exp\left\{n\sum_{j=1}^nq_j p_j (H_j^d(1)-1)\right\}
\]
for $H_i^d(1)$ for all $i$, which is exactly the same system as for the
probability $P_{ep}$ for an epidemic to occur due to failure of a node. As
the right hand side does not depend on $i$, we find that all $H_i^d$ are
equal and we will thus drop the index and simply use $H$ for all $H_i^d(1)$ in
the following calculation.

Thus, our task is solving the equation
\begin{equation}\label{eq:ER-system}
H=\exp\left\{n\sum_{j=1}^nq_j p_j (H-1)\right\}
\end{equation}
for $H$. This identity always admits the trivial solution $H=1$, which
corresponds to the case where we have a finite outbreak with probability one
and correspondingly the probability of an epidemic is zero. If the condition
in Example \ref{exa:ER-condition} is not fulfilled, i.e. if $s:=n\sum_{i=1}^n
q_i p_i
>1$, there also exists a unique solution $H(s)$ in the open interval $(0,1)$.
This solution is given by
\[
H(s)= - \frac{W(-se^{-s})}{s}
\]
where $W(z)$ denotes the principal branch of the Lambert W-function
\cite{Corless1996}.

Hence, we obtain
\[
P_{ep}=f=1-\exp\left\{s (H-1)\right\}=1-H=1+ \frac{W(-se^{-s})}{s}.
\]
It must be remarked that the equality of $f$ and $P_{ep}$ is \emph{not
generally valid} in arbitrary directed networks. Figure
\ref{fig:percol_epidemic} shows the increase in probability of epidemics (and
correspondingly in fraction of affected nodes) in dependence of the sum $s$.
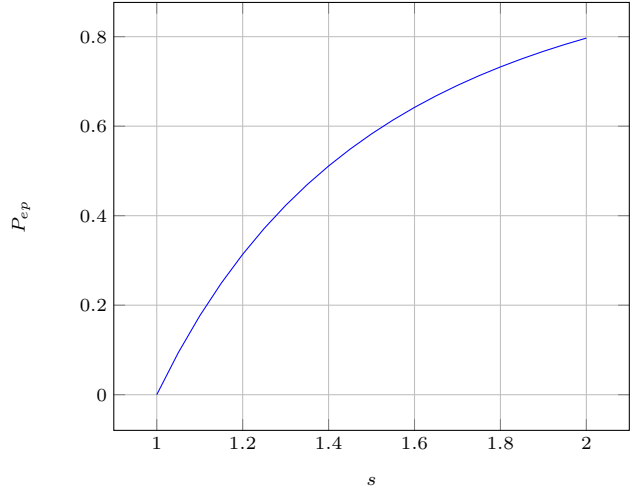
\begin{figure}
\centering
\scriptsize
\begin{tikzpicture}
\begin{axis}[xlabel=$s$, ylabel=$P_{ep}$, no markers, xmajorgrids,ymajorgrids,xminorgrids,yminorgrids]
\addplot coordinates {
 (1.,  0.)
 (1.05,  0.0937018)
 (1.1,  0.176134)
 (1.15,  0.249002)
 (1.2,  0.313698)
 (1.25,  0.37137)
 (1.3,  0.42297)
 (1.35,  0.469294)
 (1.4,  0.511011)
 (1.45,  0.54869)
 (1.5,  0.582812)
 (1.55,  0.61379)
 (1.6,  0.641981)
 (1.65,  0.667691)
 (1.7,  0.691186)
 (1.75,  0.712698)
 (1.8,  0.73243)
 (1.85,  0.75056)
 (1.9,  0.767244)
 (1.95,  0.78262)
 (2.,  0.796812)
};
\end{axis}
\end{tikzpicture}\\
\caption{Probability of an epidemic depending on the excess
$s$ of condition \eqref{eqn:finite-outbreak-criterion}}\label{fig:percol_epidemic}
\end{figure}

\end{exa}

An epidemic can as well be caused by failure of an edge of type $i$. This situation can be reduced to failure of a node with the following standard trick: replace the considered edge $u\To v$ of type $i$ by two edges  $u\To z$ and  $z\To v$ of the same type. A failure of the original edge can then be viewed as failure of the new node $z$ and the formalism derived above can be applied.

\section{Conclusion}
Percolation appears as a quite powerful technique to describe error
propagation through a network with an arbitrary topology. We replaced the
standard assumption of uniform failure by different probabilities of failure
in order to allow for different types of edges. For this scenario we found
linear equation systems that yield the expected size of an outbreak as well
as the probability of an epidemic. If the network is described by the
Erd\H{o}s-R\'enyi model, we give explicit solutions to these equation
systems.

\section{Outlook}
 While this model captures the diverse nature of components of a real life network, other aspects are still missing. Next steps could for example be introduction of countermeasures to reduce the spreading. Future work will also include construction of algorithms for simulation.

\section*{Acknowledgment}
This work was supported by the European Commission's Project No. 608090, HyRiM (Hybrid Risk Management for Utility Networks) under the 7th Framework Programme (FP7-SEC-2013-1).

\bibliographystyle{IEEEtran}
\bibliography{literature}

\begin{thebibliography}{10}
\providecommand{\url}[1]{#1}
\csname url@samestyle\endcsname
\providecommand{\newblock}{\relax}
\providecommand{\bibinfo}[2]{#2}
\providecommand{\BIBentrySTDinterwordspacing}{\spaceskip=0pt\relax}
\providecommand{\BIBentryALTinterwordstretchfactor}{4}
\providecommand{\BIBentryALTinterwordspacing}{\spaceskip=\fontdimen2\font plus
\BIBentryALTinterwordstretchfactor\fontdimen3\font minus
  \fontdimen4\font\relax}
\providecommand{\BIBforeignlanguage}[2]{{%
\expandafter\ifx\csname l@#1\endcsname\relax
\typeout{** WARNING: IEEEtran.bst: No hyphenation pattern has been}%
\typeout{** loaded for the language `#1'. Using the pattern for}%
\typeout{** the default language instead.}%
\else
\language=\csname l@#1\endcsname
\fi
#2}}
\providecommand{\BIBdecl}{\relax}
\BIBdecl

\bibitem{Poggi2013}
S.~Poggi, F.~Neri, V.~Deytieux, A.~Bates, W.~Otten, C.~Gilligan, and D.~Bailey,
  ``Percolation-based risk index for pathogen invasion: Application to
  soilborne disease in propagation systems,'' vol. 103, no.~10, pp. 1012--9,
  2013.

\bibitem{Duffie2009}
D.~Duffie, S.~Malamud, and G.~Manso, ``Information percolation with equilibrium
  search dynamics,'' vol.~77, no.~5, pp. 1513--1574, September 2009.

\bibitem{Duffie2011}
------, ``Information percolation in segmented markets,'' National Bureau of
  Economic Research, Tech. Rep., 2011, working Paper 17295.

\bibitem{Grimmett1989}
G.~Grimmett, \emph{Percolation}.\hskip 1em plus 0.5em minus 0.4em\relax
  Springer, 1989.

\bibitem{Newman2002}
M.~E.~J. Newman, ``The spread of epidemic disease on networks,'' \emph{Physical
  Review E}, vol. 66, 016128, 2002.

\bibitem{Schwartz2002}
N.~Schwartz, R.~Cohen, D.~ben Avraham, A.-L. Barabasi, and S.~Havlin,
  ``Percolation in directed scale-free networks,'' \emph{Physical Review E},
  vol. 66, 015104, 2002.

\bibitem{Cohen2002}
R.~Cohen, D.~ben Avraham, and S.~Havlin, ``Percolation critical exponents in
  scalel-free networks,'' \emph{Physical Review E}, vol. 66, 036113, 2002.

\bibitem{Sander2002}
L.~M. Sander, C.~P. Warren, I.~M. Sokolov, S.~C., and J.~Koopman, ``Percolation
  on heterogeneous networks as a model for epidemics,'' \emph{Mathematical
  Bioosciences}, vol. 180, pp. 293--305, 2002.

\bibitem{Miller2008}
J.~C. Miller, ``Bounding the size and probability of epidemics on networks,''
  \emph{Applied Probability Trust}, vol.~45, pp. 498--512, 2008.

\bibitem{Restrepo2008}
J.~G. Restrepo, E.~Ott, and B.~R. Hunt, ``Resilience of the internet to random
  breakdowns,'' \emph{Physical Review Letters}, vol.~5, no. 100, 2008.

\bibitem{Meyers2006}
L.~A. Meyers, M.~E.~J. Newman, and B.~Pourbohloul, ``Predicting epidemics on
  directed contact networks,'' \emph{Journal of Theoretical Biology}, vol. 240,
  no.~3, pp. 400--418, 2006.

\bibitem{Wilf1994}
H.~S. Wilf, \emph{Generatingfunctionology}.\hskip 1em plus 0.5em minus
  0.4em\relax Academic Press, 1994.

\bibitem{Newman2001}
M.~E.~J. Newman, S.~H. Strogatz, and D.~J. Watts, ``Random graphs with
  arbitrary degree distributions and their applications,'' \emph{Physical
  Review E}, vol. 64, 026118, 2001.

\bibitem{Cohen2000}
R.~Cohen, K.~Erez, D.~ben Avraham, and S.~Havlin, ``Resilience of the internet
  to random breakdowns,'' \emph{Physical Review Letters}, vol.~85, no.~21,
  2000.

\bibitem{Cohen2001}
------, ``Breakdown of the internet under intentional attack,'' \emph{Physical
  Review Letters}, vol.~86, pp. 3682--3685.

\bibitem{Callaway2000}
D.~S. Callaway, M.~E.~J. Newman, S.~H. Strogatz, and D.~J. Watts, ``Network
  robustness and fragility: Percolation on random graphs,'' \emph{Physical
  Review Letters}, vol.~85, no.~25, 2000.

\bibitem{Newman2011}
M.~E.~J. Newman and C.~R. Ferrario, ``Competing epidemics on complex
  networks,'' \emph{Physical Review E}, vol. 84, 036106, 2011.

\bibitem{Newman2013}
------, ``Interacting epidemics and coinfection on contact networks,''
  \emph{PLoS ONE}, vol.~8, no.~8, 2013.

\bibitem{Erdos1959}
P.~Erd\H{o}s and A.~R\'enyi, ``On random graphs,'' \emph{Publicationes
  Mathematicae}, vol.~6, pp. 290--297, 1959.

\bibitem{Cohen2010}
R.~Cohen and S.~Havlin, \emph{Complex Networks - Structure, Robustness and
  Function}.\hskip 1em plus 0.5em minus 0.4em\relax Cambridge University Press,
  2010.

\bibitem{Corless1996}
R.~M. Corless, G.~H. Gonnet, D.~E.~G. Hare, D.~J. Jeffrey, and D.~E. Knuth,
  ``On the {L}ambert {W} function,'' \emph{Computational Mathematics}, vol.~5,
  pp. 329--359, 1996.

\end{thebibliography}

\end{document}